\documentclass[twocolumn,superscriptaddress]{revtex4}
\usepackage{graphicx}
\usepackage{dcolumn}
\usepackage{bm}
\usepackage{amsmath}
\usepackage{amssymb}
\usepackage{makecell}
\usepackage{hyperref}
\hypersetup{
  colorlinks=true,
  linkcolor=blue,
  citecolor=cyan,
  urlcolor=blue
}

\newcommand{\bq}{\begin{equation}}
\newcommand{\eq}{\end{equation}}
\newcommand{\bqn}{\begin{eqnarray}}
\newcommand{\eqn}{\end{eqnarray}}

\newcommand{\lb}{\label}

\begin{document}

\title{Constraints on Schwarzschild Black Hole in a Generalized Dehnen-Type $(1,4,\gamma)$ Dark Matter Halo via the S2 Star Orbit around Sgr A$^\star$}

\author{Tursunali Xamidov}
\email{xamidovtursunali@gmail.com}
\affiliation{Institute for Theoretical Physics and Cosmology, Zhejiang University of Technology, Hangzhou 310023, China}
\affiliation{Institute of Fundamental and Applied Research, National Research University TIIAME, Kori Niyoziy 39, Tashkent 100000, Uzbekistan}
\affiliation{Tashkent State Technical University, 100095 Tashkent, Uzbekistan}

\author{Sanjar Shaymatov}
\email{sanjar@astrin.uz}
\affiliation{Institute of Fundamental and Applied Research, National Research University TIIAME, Kori Niyoziy 39, Tashkent 100000, Uzbekistan}
 \affiliation{University of Tashkent for Applied Sciences, Str. Gavhar 1, Tashkent 100149, Uzbekistan}
 \affiliation{National University of Uzbekistan, Tashkent, 100174, Uzbekistan}

\author{Qiang Wu}
\email[Corresponding author: ]{wuq@zjut.edu.cn}
\affiliation{Institute for Theoretical Physics and Cosmology, Zhejiang University of Technology, Hangzhou 310023, China}
\affiliation{United Center for Gravitational Wave Physics (UCGWP), Zhejiang University of Technology, Hangzhou, 310023, China}

\begin{abstract}	
The distribution of dark matter (DM) halo around supermassive black holes (BHs) may leave observable imprints on stellar dynamics near galactic centers. Motivated by this, we investigate the orbital motion of the S2 star in the spacetime of a recently derived generalized Schwarzschild BH solution embedded in a Dehnen-type $(1,4,\gamma)$ DM halo, considering it as a possible model for Sgr A$^{\star}$ at the center of the Milky Way. Unlike previous studies restricted to specific values of the halo parameter $\gamma$, the present solution describes the fully generalized case with arbitrary $\gamma$. We derive the corresponding equations of motion and obtain the associated pericenter shift over one orbital period. Using observational data of the S2 star, we constrain the parameters of the Schwarzschild--Dehnen BH-DM system through a Markov Chain Monte Carlo (MCMC) analysis. 
Our MCMC analysis provides constraints on the parameter $\gamma$ of the generalized Dehnen-type DM halo. The posterior medians of $\gamma$ with 68\% credible intervals are $\gamma=0.652^{+0.435}_{-0.446}$ and
$\gamma=0.771^{+0.486}_{-0.522}$
for Dataset~1 from Do et al.~\cite{Do19} and Dataset~2 from
Gillessen et al.~\cite{Gillessen17ApJ}, respectively.
The corresponding one-sided 95\% upper credible limits are $\gamma<1.238$ and $\gamma<1.389$. 
If the halo is instead normalized to the observed local dark-matter density at solar distance, $\rho_{\rm DM}(8\, \rm kpc)\simeq0.01\,M_\odot\,{\rm pc}^{-3}$ \cite{McMillan2017},
the 95\% upper limits become $\gamma<1.949$ and $\gamma<1.931$ for Dataset 1
and Dataset 2, respectively.
The analysis shows that the S2 orbital motion provides constraints on the parameter $\gamma$ of the DM halo profile, while the characteristic halo density $\rho_s$ and scale radius $r_s$ exhibit a strong degeneracy and remain sensitive to the adopted prior ranges.

\end{abstract}

\maketitle

\section{\label{sec:intro}Introduction}

General Relativity (GR), developed by Albert Einstein in 1915, describes gravity as the curvature of space-time caused by mass and energy~\cite{Einstein1916}. The theory successfully explains several observational phenomena, including gravitational lensing, the pericenter precession of Mercury, and gravitational redshift. Over the past decades, numerous experiments and observations have tested General Relativity (GR) with remarkable precision. One of the earliest major confirmations was provided by the binary pulsar PSR B1913+16, whose orbital decay matched the energy loss predicted from gravitational-wave emission~\cite{Hulse74,Hulse75}. More recently, GR has been further validated through direct gravitational wave (GW) detections by the LIGO–Virgo collaboration~\cite{Abbott16a,Abbott16b}, observations of stellar orbits around the supermassive BH at the Galactic center~\cite{Ghez:1998ph}, and the imaging of the BH shadow in M87 by the Event Horizon Telescope (EHT)~\cite{Akiyama19L1,Akiyama19L6,Akiyama22L12}. To date, no experimentally confirmed deviation from the predictions of GR has been observed, highlighting its extraordinary success as a theory of gravitation.

Despite its remarkable success, it remains unclear whether GR is complete as it does not describe spacetime singularities or explain dark matter and dark energy. Furthermore, the observed accelerated expansion of the Universe~\cite{Di_Valentino_2021} may indicate the need for modifications to GR on cosmological scales, potentially motivating new alternative theories and models of gravity \cite{Peebles03,Spergel07,Wetterich88,Caldwell09,Kiselev03,Rubin80,Persic96}. Future observations of BHs and GWs will provide even more strict tests of the theory. Therefore, constructing gravitational solutions describing BH spacetimes embedded in DM environments is of considerable importance, while the fundamental nature of DM still remains unknown in modern cosmology and GR.

Current astrophysical observations strongly support the existence of dark matter (DM), which constitutes roughly 90\% of the total galactic mass, with luminous baryonic matter contributing the remaining 10\% \cite{Persic96}. Observations further reveal that many massive spiral and elliptical galaxies contain central supermassive black holes (BHs) embedded within surrounding DM halos \cite{Valluri04ApJ,Akiyama19L1,Akiyama19L6,Akiyama22L12}. A variety of theoretical models extending the Standard Model predict the existence of weakly interacting DM particle candidates, including WIMPs, axions, and neutrinos \cite{Boehm04NPB,Bertone05PhR,Feng09JCAP,Schumann19}. Owing to their extremely weak interactions with ordinary matter, the existence and properties of these particles are primarily inferred through gravitational effects. Consequently, DM is expected to accumulate in the vicinity of supermassive BHs, where it may significantly influence both extreme \cite{Babak17PRD} and intermediate \cite{Brown07PRL,Amaro-Seoane18PRD} mass ratio inspirals, thereby providing a potential probe of DM halo profiles. Furthermore, DM halos are essential for explaining galactic rotation curves and the dynamics of galaxy clusters \cite{Rubin70ApJ,Bertone18Nature,Corbelli00MNRAS,Clowe06ApJL}. BHs embedded in DM halos provide a natural framework for studying BH–DM interactions and probing the nature of DM. Several analytical halo models, including the Einasto, Navarro–Frenk–White, Burkert, and Dehnen-type profiles, have been considered in this context~\cite{Dutton_2014,Navarro96ApJ,Burkert95ApJ,Dehnen93,Shukirgaliyev21A&A,Pantig22JCAP,Al-Badawi25JCAP,Wang2025PhRvD,Li-Yang12,Shaymatov21d,Shaymatov21pdu,Shaymatov22a,Cardoso22DM,Hou18-dm,Shen24PLB,Shen25PLB}. The surrounding DM distribution can modify the spacetime geometry and affect observable quantities such as the ISCO, BH shadows, quasinormal modes, and GW signatures~\cite{Gohain24DM,Uktamov25EPJC,Al-Badawi25CPC,Al-Badawi25CTP_DM,Alloqulov-Xamidov25,BoWang2025JCAP...01..086L,Xamidov25PDU,Xamidov25EPJC...85.1193X}.

It must be emphasized that the Galactic center of Sgr A$^{\star}$ provides an important strong-gravity laboratory for testing BH spacetimes and modified theories of gravity. Observations of the S-cluster stars constrain the mass of Sgr A$^{\star}$ to $M \simeq 4.299\times10^6 \,M_\odot$ at a distance of $\sim 8.276    \,\mathrm{kpc}$~\cite{Gravity2024A&A692A242G}. Despite significant observational progress, the physical properties of its accretion environment, including the possible presence of a jet or accretion disk, remain uncertain. Future high-resolution observations of the BH shadow are expected to provide tighter constraints on the mass, spin, and near-horizon geometry of Sgr A$^{\star}$~\cite{Perlick21}. Recent analyses have shown that the S2 star orbit can be used to test alternative models of Sgr A$^{\star}$, including wormholes~\cite{Jusufi21}, DM spike models~\cite{Nampalliwar21ApJ}, and loop quantum gravity inspired spacetimes~\cite{Yan22JCAP}. {
The S2 orbit and the black hole shadow have also been used to constrain a generalized double power-law dark matter halo around Sgr A$^\star$ \cite{ZhenLiAA2026}.
} Furthermore, stellar orbital observations at the Galactic center were used to test quantum-corrected Schwarzschild spacetime in loop quantum gravity~\cite{WuQiang2023PhRvD} and the spherically symmetric parameterized Rezzolla-Zhidenko spacetime \cite{Shaymatov2023ApJ}. 
 
Motivated by these studies, in this work, we analyze the orbital motion of the S2 star in a Schwarzschild-like BH spacetime embedded in a Dehnen-type DM halo with density profile $(1,4,\gamma)$, considering it as a possible description of the Sgr A$^{\star}$ at the center of the Milky Way. We derive the corresponding equations of motion and compute the pericenter precession of test particle orbits. We further constrain the model parameters using the orbital motion of the S2 star around Sgr A$^{\star}$ \cite{Lacroix18,Nucita07,Ghez05,Ghez00} by comparing the theoretical predictions with the observational data of Do et al.~\cite{Do19} and Gillessen et al.~\cite{Gillessen17ApJ}.

The structure of the paper is as follows: In
Sec.~\ref{Sec:II}, we briefly describe the spacetime metric and derive the corresponding equations of motion for time-like test
particle geodesics moving around the Schwarzschild-like BH in the Dehnen-type DM halo with density profile $(1,4,\gamma)$, which is followed by the main derivation of the expression for the pericenter shift. In Sec.~\ref{Sec:III}, we obtain constrains on the model parameters using the orbital motion and data of the S2 star around Sgr A$^{\star}$ and determine the {posterior median} values and corresponding upper limits of the model parameters by employing Markov Chain Monte Carlo (MCMC) method. Sec.~\ref{conclusion} is devoted to our concluding remarks. Throughout the paper, we use a system of units $G=c=1$ and the spacetime signature
$(-,+,+,+)$.

\section{\label{Sec:II}
	Spacetime metric and timelike geodesics}

Within a spherically symmetric spacetime, the mass distribution is determined by the dark matter density profile, and the corresponding mass profile is given by
\begin{eqnarray}
M_{\rm DM}(r)&=&4\pi\int^{r}_{0}\rho_{\rm DM}(r_1)\, r_1^2\,dr_1 \, .
\end{eqnarray}
where $\rho_{\rm DM}(r_1)$ denotes the density profile. {The general double power-law density profile introduced by  \cite{Zhao_1996MNRAS278_488Z} (see  \cite{Mo10book})} is given by 
\begin{eqnarray}\label{eq.density}
\rho_{\rm DM}(r_1)=\rho_s\left(\frac{r_1}{r_s}\right)^{-\gamma}
\left[1+\left(\frac{r_1}{r_s}\right)^\alpha\right]^{\frac{\gamma-\beta}{\alpha}} \, .
\end{eqnarray}
Here, $\rho_s$ and $r_s$ are the characteristic density and scale radius, respectively, while $\alpha$, $\beta$, and $\gamma$ control the shape of
the density profile. It is worth noting that for $(\alpha,\beta,\gamma)=(1,4,\gamma)$, the general density profile given in Eq.~\eqref{eq.density} reduces to the Dehnen density profile. The parameter $\gamma$ lies in the range $[0,3)$. 
{Recently, Boltaev et al.~\cite{boltaev2026arxiv} derived the fully generalized Schwarzschild-like black hole solution surrounded by a Dehnen-type dark matter halo, extending previous solutions restricted to specific values of $\gamma$ (see. e.g., \cite{Gohain24DM,Al-Badawi25JCAP,Uktamov25EPJC})}. 
The previous studies considered particular cases of the Dehnen profile, with $\gamma=0$ in Ref.~\cite{Gohain24DM}, $\gamma=2$ in Ref.~\cite{Uktamov25EPJC}, and $\gamma=5/2$ in
Ref.~\cite{Al-Badawi25JCAP}, where the corresponding DM metric functions
were constructed using the tangential-velocity relation. In contrast,
the generalized solution of Ref.~\cite{boltaev2026arxiv} was obtained
directly from the Einstein field equations for arbitrary $\gamma$.

The corresponding metric can be written as follows:
\begin{eqnarray}\label{eq.full-line}
    ds^2=-f(r)dt^2+\frac{1}{f(r)}dr^2+r^{2}\left(
d\theta ^{2}+\sin ^{2}\theta d\phi ^{2}\right)\, , 
\end{eqnarray}
where 
\begin{eqnarray}\label{Eq.radial_func}
    f(r)=1-\frac{2M}{r}-\frac{8\pi\rho_sr_s^3}{(3-\gamma)r}\Big(\frac{r}{r+r_s}\Big)^{3-\gamma}\, .
\end{eqnarray}

Eq.~\eqref{Eq.radial_func} follows directly from the Einstein field equations
for the profile in Eq.~\eqref{eq.density} with $(\alpha,\beta)=(1,4)$ and the
condition $P_r=-\rho_{\rm DM}$, which implies $A(r)=B(r)=f(r)$ for the metric
$ds^2=-A(r)\,dt^2+dr^2/B(r)+r^2d\Omega^2$. The $tt$ component of the field
equations gives (see \cite{boltaev2026arxiv} for details)
\begin{equation}
\frac{d}{dr}\Big[r\big(B(r)-1\big)\Big]=-8\pi r^2\rho_{\rm DM}(r),
\end{equation}
so that
\begin{equation}
f(r)=B(r)=1-\frac{2M}{r}-\frac{2M_{\rm DM}(r)}{r},
\end{equation}
with
\begin{equation}
M_{\rm DM}(r)=\frac{4\pi\rho_s r_s^3}{3-\gamma}
\left(\frac{r}{r+r_s}\right)^{3-\gamma}\, .
\end{equation}
For comparison with Ref.~\cite{Al-Badawi25JCAP}, setting $\gamma=5/2$ in
Eq.~\eqref{Eq.radial_func} gives
\begin{equation}
f(r)=1-\frac{2M}{r}
-\frac{16\pi\rho_s r_s^3}{r}
\sqrt{\frac{r}{r+r_s}}.
\end{equation}
This expression differs from the metric function of
Ref.~\cite{Al-Badawi25JCAP} obtained from the tangential-velocity relation,
\begin{equation}
A_{\rm AB}(r)=\exp\left[-32\pi\rho_s r_s^2\sqrt{1+\frac{r_s}{r}}\right].
\end{equation}
To ensure asymptotic flatness, $A_{\rm AB}(\infty)=1$, the above expression should be rewritten in the normalized form
\begin{equation}
A_{\rm AB}^{\rm norm}(r)=
\exp\left[
-32\pi\rho_s r_s^2
\left(
\sqrt{1+\frac{r_s}{r}}-1
\right)
\right].
\end{equation}
In the far-field limit, $r\gg r_s$, and after including the Schwarzschild contribution, both metric functions then show the same asymptotic behavior,
\begin{equation}
f(r)\simeq
1-\frac{2M}{r}
-\frac{16\pi\rho_s r_s^3}{r}
+\mathcal{O}(r^{-2}).
\end{equation}
Thus, the two constructions agree at leading order in the far-field limit,
while their difference at finite radii arises from the different construction
procedures. The solution obtained by ~\cite{boltaev2026arxiv} is also independently supported by the recent work
of Bolokhov~\cite{Bolokhov2026EPJC}.

We investigate the impact of a dark matter halo on spacetime by analyzing the dynamics of massive particles orbiting a Schwarzschild-like black hole surrounded by a Dehnen-type DM halo. We further constrain the DM halo parameters $\gamma$, $\rho_s$ and $r_s$ using S2 star data. To describe the motion of a particle in the spacetime of a Schwarzschild-like black hole characterized by the metric \eqref{eq.full-line}, we employ the Hamiltonian formalism \cite{Misner73}. The Hamiltonian is expressed as
\begin{equation}\label{hamiltonian}
H=\frac{1}{2}g^{\mu\nu}p_\mu p_\nu \ ,
\end{equation}
where $p^\mu=m\,u^\mu$ and $u^\mu$ denote the four-momentum and four-velocity, respectively. In spherical coordinates, the indices $\mu$ and $\nu$ correspond to $(t, r, \theta, \phi)$. For massive particles, the Hamiltonian satisfies $H=-m^2/2$. The equations of motion are then derived from Hamilton’s equations. The particle dynamics are governed by Hamilton’s equations,
\begin{equation} \label{Hamx}
\frac{dx^\mu}{d\lambda} = \frac{\partial H}{\partial p_\mu} \,, \qquad
\frac{dp_\mu}{d\lambda} = -\frac{\partial H}{\partial x^\mu} \, ,
\end{equation}
where $\lambda = \tau/m$ is the affine parameter and $\tau$ denotes the proper time. Since the spacetime metric \eqref{eq.full-line} is independent of the coordinates $t$ and $\phi$, the particle's energy and angular momentum are conserved, with $p_t=-E$ and $p_\phi=L$ \cite{Misner73}. By employing Eqs.~\eqref{hamiltonian} and \eqref{Hamx}, the equations governing the radial and azimuthal motion of the particle in the equatorial plane can be written as
\begin{align} \label{radial}
\left(\frac{dr}{d\tau}\right)^2 
= \mathcal{E}^2 - f(r)\left(1 + \frac{\mathcal{L}^2}{r^2}\right) \,, 
\quad 
\frac{d\phi}{d\tau} = \frac{\mathcal{L}}{r^2} \, ,
\end{align}
where $\mathcal{E} = E/m$ and $\mathcal{L} = L/m$ are the specific energy and specific angular momentum, respectively. From Eq.~\eqref{radial}, the effective potential can be expressed as
\begin{eqnarray}\label{Veff_general}
    V_{\rm eff}(r)&=&\left[1-\frac{2M}{r}-\frac{8\pi\rho_sr_s^3}{(3-\gamma)r}\Big(\frac{r}{r+r_s}\Big)^{3-\gamma}\right] \nonumber\\
    &\times&
    \left(1+\frac{\mathcal{L}^2}{r^2}\right)\, .
\end{eqnarray}
From Eqs.~\eqref{radial} and \eqref{Veff_general}, the trajectory equation is obtained as
\begin{eqnarray} \label{trajectory-eqn}
    \left(\frac{dr}{d\phi}\right)^2 
    = \frac{r^4\left(  \mathcal{E}^2 
      - V_{\rm eff}(r)\right)}{\mathcal{L}^2}  \, .
\end{eqnarray}
After introducing the transformation $u=1/r$ and differentiating with respect to $\phi$, we obtain the following equation:
\begin{eqnarray}\label{geod-eq}
    \frac{d^2 u}{d\phi^2} =\frac{M}{\mathcal{L}^2} - u + \frac{g(u)}{\mathcal{L}^2} \, ,
\end{eqnarray}
where 
\begin{eqnarray}
    \frac{g(u)}{\mathcal{L}^2}&=&3 M u^2-4 \pi  \rho_s r_s^3 \big(1+r_s u\big)^{\gamma -4}\nonumber\\
    &\times&\left(\frac{1+ \gamma  \mathcal{L}^2 r_s u^3+3 \mathcal{L}^2 u^2+\gamma  r_s u-2 r_s u}{(\gamma -3) \mathcal{L}^2}\right) \, .
\end{eqnarray}
Using the approach of Ref.~\cite{Adkins_2007,boltaev2026arxiv}, the pericenter shift per orbital period is
\begin{equation} \label{p-shift-eqn}
    \Delta \phi = \frac{\pi}{\mathcal{L}^2} \left| \frac{d g(u)}{du} \right|_{u = \frac{1}{b}}\, ,
\end{equation}
where $b=a(1-e^2)$, with $a$ and $e$ denoting the semi-major axis and the orbital eccentricity, respectively. To recover physical dimensions, $M$, $L^2$, $\rho_s$ and $r_s$ are rescaled as
\begin{eqnarray}
    \left\{
\begin{aligned}
M  &\Rightarrow \frac{GM}{c^2} \, ,\\
L^2 &\Rightarrow \frac{GMa(1 - e^2)}{c^2}\, ,
\end{aligned}
\right.\qquad 
\quad
\left\{
\begin{aligned}
\rho_s &\Rightarrow \frac{3c^4}{32\pi G^2M^2}\,\rho_s \, ,\\
r_s &\Rightarrow \frac{GM}{c^2}\,r_s\, .
\end{aligned}
\right.
\end{eqnarray}
Here, the quantities $\rho_s$ and $r_s$ on the right-hand side are dimensionless. From the above equations together with Eq.~\eqref{p-shift-eqn}, the pericenter shift after one complete revolution is obtained as
\begin{widetext}
\begin{eqnarray}\label{p.shift2}
	\triangle\phi=6 \pi  \alpha + \frac{3 \pi  \alpha \rho_s r_s^3(1+\alpha r_s)^{\gamma-4}}{8 (\gamma -3)} \left(2 r_s-\gamma  r_s(1+3 \alpha)-6-\frac{ (\gamma -4)r_s \left(1+\alpha ((\gamma -2) r_s+3)+\alpha^2 \gamma  r_s\right)}{ 1+\alpha r_s}\right)\, ,
\end{eqnarray}
\end{widetext}
where
\begin{equation}
\alpha = \frac{G M}{a c^2 (1 - e^2)}.
\end{equation}

\section{Constraints on parameters of Schwarzschild BH immersed in DM halo via the S2 star orbit data \label{Sec:III} }

The center of our galaxy provides a valuable natural laboratory for studying the behavior of black holes (BHs) and testing theories of modified gravity under extreme physical conditions. At the galactic center, the black hole Sgr A$^\star$ is surrounded by a group of stars known as the S clusters. Detailed observational data on their astronomical positions are publicly available, and from this information, constraints on the black hole's parameters can be derived \cite{Gillessen17ApJ, Do19, GRAVITY2}. 
Specifically, we use two independent datasets: Dataset 1 from
Do et al.~\cite{Do19} and Dataset 2 from Gillessen et al.~\cite{Gillessen17ApJ}. For each dataset, we use the astrometric positions and radial-velocity measurements of the S2 star. The details of the datasets are summarized
below.

\begin{enumerate}
    \item \textbf{Dataset 1}:
    \begin{itemize}
        \item \textbf{Astrometric Positions}: A total of 45 astrometric positions, spanning from 1995.439 to 2018.394, as reported in \cite{Do19}. Among these, 11 new astrometric measurements from 2016 to 2018 are included in our analysis. The data for these positions were gathered using speckle imaging (1995-2005) and adaptive optics (AO) imaging (2005-2018) at the W. M. Keck Observatory.
        \item \textbf{Radial Velocities}: A total of 115 radial velocity measurements between 2000.476 and 2018.707, as reported in \cite{Do19}. These measurements were obtained from various telescopes, including the W. M. Keck Observatory, Gemini North Telescope, and Subaru Telescope. All observations were taken using adaptive optics (AO).
    \end{itemize}
    \item \textbf{Dataset 2}:
    \begin{itemize}
        \item \textbf{Astrometric Positions}: A total of 145 astrometric positions of the S2 star, spanning from 1992.224 to 2016.53, as reported in \cite{Gillessen17ApJ}. Data before 2002 were observed using the ESO New Technology Telescope (NTT), and data after 2002 were obtained from the Very Large Telescope (VLT).
        \item \textbf{Radial Velocities}: 44 radial velocity measurements between 2000.487 and 2016.519 as reported in \cite{Gillessen17ApJ}. Data before 2003 were collected from NIRC2, and data after 2003 were obtained from the INtegral Field Observations in the Near Infrared (SINFONI).
    \end{itemize}

\end{enumerate}
Using the above datasets, we constrain the model parameters through a Markov Chain Monte Carlo (MCMC) analysis implemented with the \textit{emcee} Python package \cite{Foreman_Mackey_2013}.

We now turn to the study of the S2 star orbit using the equations of motion for massive particles around compact objects. To this end, we first numerically solve Eqs. \eqref{Hamx} and \eqref{radial} with initial conditions  ${t(\lambda_0), r(\lambda_0), \phi(\lambda_0)}$ and their derivatives ${\dot{t}(\lambda_0), \dot{r}(\lambda_0), \dot{\phi}(\lambda_0)}$. This allows us to describe the motion of the S2 star in the orbital plane. Since the observational data are given in the celestial plane, a transformation between the orbital and celestial planes is required for a direct comparison. The transformation from the orbital plane $(x,y,z)$ to the celestial plane $(X,Y,Z)$ is given by
\begin{eqnarray}
	X &=& x B+y G\label{Xobs}\, ,\\
	Y&=&xA+yF\label{Yobs}\, ,\\
	Z&=&xC+yH \label{Zobs}\, .
\end{eqnarray}
Here the coefficients $A,B,C,F,G,H$ are defined as
\begin{eqnarray}
	A&=&\cos \Omega \cos \omega -\sin \Omega \sin \omega \cos \iota\, . \lb{A}\\
	B&=&\sin \Omega \cos \omega +\cos \Omega \sin\omega \cos \iota\, , \lb{B}\\
	C&=&\sin \omega \sin \iota\, ,  \lb{C}\\
	F&=&-\cos\Omega \sin \omega -\sin \Omega \cos \omega \cos \iota\, , \lb{F}\\
	G&=&-\sin \Omega \sin \omega +\cos \Omega \cos \omega \cos \iota\, ,  \lb{G}\\
	H&=&\cos \omega \sin \iota\, , \lb{H}
\end{eqnarray}
where $\omega$, $\Omega$ and $\iota$ represent the argument of pericenter, the longitude of the ascending node, and the orbital inclination of the S2 star, respectively.
Since positional offsets and linear drifts may arise between the gravitational center and the astrometric reference frame, we introduce the parameters $x_0$, $y_0$, $v_{x0}$, and $v_{y0}$ (see, e.g., Do et al.~\cite{Do19})
\begin{eqnarray}
	X_{\rm obs}(t_{\rm obs})&=&X(t_{\rm em})+x_{0}+v_{\rm x0}(t_{\rm obs}-t_{\rm ref})\, ,  \lb{x0} \\
	Y_{\rm obs}(t_{\rm obs})&=&Y(t_{\rm em})+y_{0}+v_{\rm y0}(t_{\rm obs}-t_{\rm ref})\, .  \lb{y0}
\end{eqnarray}
Here, $t_{\rm ref}$ defines the reference epoch for the offset and drift parameters $x_0$, $y_0$, $v_{x0}$, and $v_{y0}$, whereas $t_{\rm em}$ corresponds to the emission epoch. The parameters $x_0$ and $y_0$ describe the positional offsets, while the terms $v_{x0}(t_{\rm obs}-t_{\rm ref})$ and $v_{y0}(t_{\rm obs}-t_{\rm ref})$ describe the linear drifts of the astrometric reference frame. In our analysis, we adopt $t_{\rm ref}=1994.3$.

\begin{table}[ht]
\centering
\renewcommand{\arraystretch}{1.18}
\setlength{\tabcolsep}{12pt}

\begin{tabular}{lc}
\hline\hline
\textbf{Parameter} & \textbf{Prior} \\
\hline

\multicolumn{2}{c}{\textit{Orbital and black hole parameters}} \\
\hline
$a$ (mas)                    & $\mathcal{U}(122,135)$ \\
$e$                          & $\mathcal{U}(0.80,0.95)$ \\
$M~(10^6M_\odot)$            & $\mathcal{U}(3.0097,5.5895)$ \\
$t_{\rm apo}$ (yr)           & $\mathcal{U}(1994.0,1994.8)$ \\
$i~({}^{\circ})$             & $\mathcal{U}(132.067,138.000)$ \\
$\omega~({}^{\circ})$        & $\mathcal{U}(63.000,68.663)$ \\
$\Omega~({}^{\circ})$        & $\mathcal{U}(220.000,235.571)$ \\
$R_0$ (kpc)                  & $\mathcal{U}(5.7931,10.7587)$ \\

\hline
\multicolumn{2}{c}{\textit{Reference-frame parameters}} \\
\hline
$x_0$ (mas)                  & $\mathcal{U}(-10,10)$ \\
$y_0$ (mas)                  & $\mathcal{U}(-6,6)$ \\
$v_{x0}$ (mas/yr)            & $\mathcal{U}(-8,8)$ \\
$v_{y0}$ (mas/yr)            & $\mathcal{U}(-5,5)$ \\
$v_{z0}$ (km/s)              & $\mathcal{U}(-50,50)$ \\

\hline
\multicolumn{2}{c}{\textit{Dark-matter halo parameters}} \\
\hline
$\gamma$                     & $\mathcal{U}(0,3)$ \\
$\rho_s~(M_\odot\,{\rm pc}^{-3})$
                             & $\mathcal{U}(0,100)$ \\
$r_s$ (kpc)                  & $\mathcal{U}(3,40)$ \\

\hline\hline
\end{tabular}
\caption{{Prior distributions adopted in the MCMC analysis. Here, $\mathcal{U}(a,b)$ denotes a uniform distribution over the interval $[a,b]$.}}
\label{tab:priors}
\end{table}

\begin{figure*}
	\centering
	\includegraphics[width=0.9\textwidth]{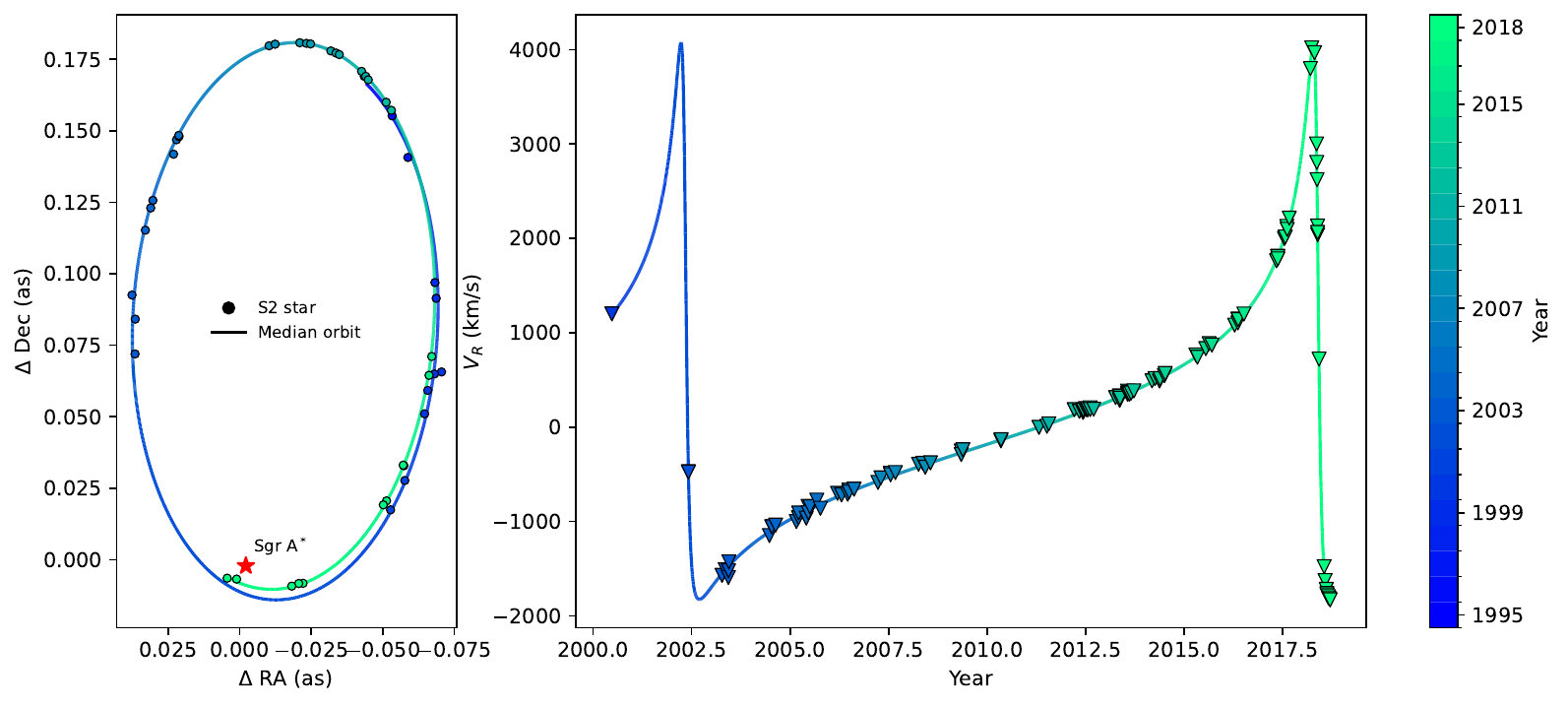}
    \includegraphics[width=0.9\textwidth]{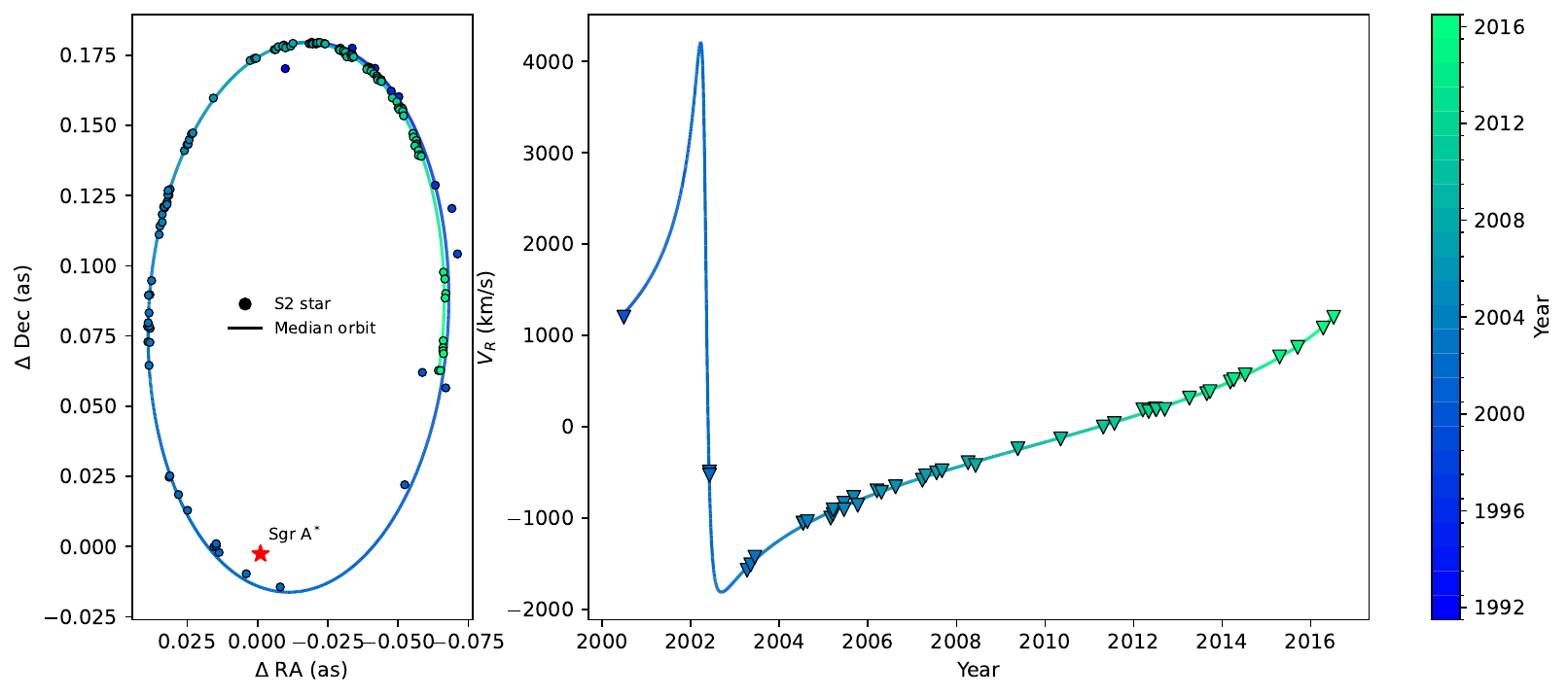}
	\caption{{
    Orbital motion and radial velocity of the S2 star around Sgr~A*. 
    The left panels show the astrometric trajectory in the plane of the sky ($\Delta$RA vs $\Delta$Dec), 
    while the right panels display the radial velocity as a function of time. 
    The top and bottom panels correspond to Dataset~1 and Dataset~2, respectively. 
    The points represent observational data, color-coded by observation year, and the solid curves indicate the posterior median model obtained from the MCMC analysis. 
    The red star marks the position of Sgr~A*. 
    }
    }  \label{fig:orbit}
\end{figure*}

Several relativistic effects must be taken into account to compare the theoretical positions with the astrometric data. We begin by considering the Roemer time delay, which modifies the arrival time of light emitted from the orbiting star depending on its position relative to the Earth. The Roemer delay is defined as
\begin{eqnarray}
	t_{\rm obs}-t_{\rm em}=\frac{Z(t_{\rm em})}{c}\, ,
\end{eqnarray}
where $t_{\rm obs}$ is the observation epoch and $Z$ is obtained from Eq.~\eqref{Zobs}. Since this equation is not straightforward to solve analytically, an iterative scheme can be employed (see for details \cite{Do19,GRAVITY1})
\begin{eqnarray}
	t_{\rm em}^{(i+1)}=t_{\rm obs}-\frac{Z(t_{\rm em}^{(i)})}{c}\, .
\end{eqnarray}
After one iteration, the equation becomes
\begin{eqnarray}
	t_{\rm em} \approx t_{\rm obs}-\frac{Z(t_{\rm obs})}{c}\, .
\end{eqnarray}

We now consider the relativistic frequency shift of the S2 star. The total observed redshift contains contributions from both the special-relativistic Doppler effect and the gravitational redshift \cite{GRAVITY2025}. For a static spherically symmetric spacetime, it can be expressed as
\begin{equation}
1+\zeta
=
\frac{1+\beta_{\rm los}}
{\sqrt{1-\beta^2}}
\frac{1}{\sqrt{-g_{tt}(r)}} ,
\end{equation}
where $\beta=v/c$ denotes the dimensionless orbital speed of the star measured by a local static observer, and $\beta_{\rm los}$ is the corresponding line-of-sight component. We also include the parameter $v_{z0}$, which accounts for the systemic line-of-sight velocity of Sgr A$^\star$. The model radial velocity is therefore given by \cite{Reid07ApJ}

\begin{eqnarray}
	V_{\rm R}=c \cdot \zeta + v_{\rm z0}\, .
\end{eqnarray}

We now proceed to perform a Markov Chain Monte Carlo (MCMC) analysis to constrain the parameters of the Schwarzschild BH embedded in Dehnen-type DM halo (see, e.g., \cite{Foreman_Mackey_2013}). The set of parameters is defined as follows:
\begin{eqnarray}
	\{M, R_{0}, a, e, i, \omega, \Omega, t_{\rm apo}, x_{0}, y_{0}, v_{x0}, v_{y0}, v_{z0},\gamma,\rho_s,r_s\}\label{paras}\, ,
\end{eqnarray}
where $M$ and $R_0$ are the black hole mass and its distance from the Earth, respectively. The parameters $\{a,e,i,\omega,\Omega,t_{\rm apo}\}$ specify the orbital configuration of the S2 star, while $\{x_0,y_0,v_{x0},v_{y0}\}$ describe the astrometric reference-frame offsets and drifts. The parameter $v_{z0}$ accounts for the systemic line-of-sight velocity. The remaining three parameters, $\{\gamma,\rho_s,r_s\}$, are associated with the Schwarzschild black hole embedded in a Dehnen-type dark matter halo, as described previously. It is important to emphasize that the orbit of the S2 star deviates from a perfect ellipse due to precession. At any given point along the trajectory, a corresponding ellipse --- referred to as the osculating ellipse --- can be defined and is characterized by the orbital elements introduced above.

In the MCMC analysis, we adopt uniform priors for the model parameters, with the complete prior ranges summarized in Table~\ref{tab:priors}. For the dark-matter halo parameters, we use broad priors
\begin{equation}
\gamma\sim\mathcal{U}(0,3);\,
\rho_s\sim\mathcal{U}(0,100)\;M_\odot\,{\rm pc}^{-3};
\,
r_s\sim\mathcal{U}(3,40)\;{\rm kpc}.
\end{equation}
The adopted range of $r_s$ is chosen to cover Galactic-scale values reported in previous studies of the Milky Way dark-matter halo ~\cite{Xiaowei_2024MNRAS, Cautun_2020MNRAS,Shen24PLB}. The halo parameters $\rho_s$ and $r_s$ are sampled in physical units and converted to geometric units, $\hat\rho_s=G^3M^2\rho_s/c^6$ and $\hat r_s=c^2r_s/(GM)$, before the geodesic equations are integrated
numerically.

The likelihood function $\mathcal{L}$ consists of two terms:
the astrometric and radial-velocity contributions,
\begin{eqnarray}
\log {\cal L} = \log {\cal L}_{\rm AP} + \log {\cal L}_{\rm VR} \, ,\label{likelyhood}
\end{eqnarray}
where the first term, $\log \mathcal{L}_{\rm AP}$, corresponds to the likelihood of the astrometric data and is written as
\begin{eqnarray}
\log {\cal L}_{\rm AP} &=& - \frac{1}{2} \sum_{i} \frac{(X_{\rm obs}^i -X_{\rm the}^i)^2}{(\sigma^i_{X,{\rm obs}})^2} \nonumber\\
&-&\frac{1}{2} \sum_{i} \frac{(Y_{\rm obs}^i -Y_{\rm the}^i)^2}{(\sigma^i_{Y,{\rm obs}})^2}\, ,
\end{eqnarray}
and the second term, $\log \mathcal{L}_{\rm VR}$, corresponds to the likelihood of the radial-velocity data and is given by
\begin{eqnarray}
\log {L}_{\rm VR} =- \frac{1}{2} \sum_{i} \frac{(V_{\rm R, obs}^i - V_{\rm R, the}^i)^2}{(\sigma^i_{V_{\rm R, obs}})^2}\, ,
\end{eqnarray}
where the subscripts “obs” and “the” refer to observational data and theoretical predictions, respectively, for the astrometric positions and radial velocities. The quantities
$\sigma^i_{X,\rm obs}$, $\sigma^i_{Y,\rm obs}$, and $\sigma^i_{V_R,\rm obs}$ denote the corresponding observational uncertainties.

\begin{table*}[t]
\centering
\renewcommand{\arraystretch}{1.22}
\setlength{\tabcolsep}{12pt}

\begin{tabular}{lcc}
\hline\hline

\textbf{Parameter} &
\makecell{\textbf{Dataset 1} \\ Do et al.~\cite{Do19}} &
\makecell{\textbf{Dataset 2} \\ Gillessen et al.~\cite{Gillessen17ApJ}} \\

\hline

\multicolumn{3}{c}{\textit{Orbital and black hole parameters}} \\
\hline

$a$ (mas)
& $126.570^{+0.263}_{-0.256}$
& $127.322^{+1.338}_{-1.266}$ \\

$e$
& $0.885695^{+0.000344}_{-0.000349}$
& $0.886708^{+0.002695}_{-0.002684}$ \\

$M~(10^6M_\odot)$
& $3.970^{+0.042}_{-0.041}$
& $4.238^{+0.223}_{-0.215}$ \\

$t_{\rm apo}$ (yr)
& $1994.3073^{+0.0021}_{-0.0021}$
& $1994.2677^{+0.0200}_{-0.0197}$ \\

$i~(^\circ)$
& $133.868^{+0.144}_{-0.146}$
& $134.167^{+0.519}_{-0.521}$ \\

$\omega~(^\circ)$
& $66.261^{+0.098}_{-0.103}$
& $65.608^{+0.709}_{-0.746}$ \\

$\Omega~(^\circ)$
& $227.787^{+0.125}_{-0.126}$
& $227.297^{+0.692}_{-0.688}$ \\

$R_0$ (kpc)
& $7.964^{+0.042}_{-0.042}$
& $8.111^{+0.213}_{-0.213}$ \\

\hline
\multicolumn{3}{c}{\textit{Reference-frame parameters}} \\
\hline

$x_0$ (mas)
& $-2.155^{+0.327}_{-0.329}$
& $-1.077^{+0.721}_{-0.797}$ \\

$y_0$ (mas)
& $-2.081^{+0.366}_{-0.358}$
& $-2.605^{+0.944}_{-0.984}$ \\

$v_{x0}$ (mas/yr)
& $0.098^{+0.015}_{-0.015}$
& $0.130^{+0.054}_{-0.049}$ \\

$v_{y0}$ (mas/yr)
& $0.231^{+0.017}_{-0.017}$
& $0.009^{+0.074}_{-0.071}$ \\

$v_{z0}$ (km/s)
& $-10.668^{+2.510}_{-2.439}$
& $11.033^{+7.721}_{-7.592}$ \\

\hline
\multicolumn{3}{c}{\textit{Dark matter halo parameters}} \\
\hline

$\gamma$
& $0.652^{+0.435}_{-0.446}$
& $0.771^{+0.486}_{-0.522}$ \\

$\rho_s~(M_\odot\,{\rm pc}^{-3})$
& $48.35^{+35.22}_{-33.79}$
& $48.08^{+35.50}_{-33.43}$ \\

$r_s~({\rm kpc})$
& $20.51^{+12.94}_{-12.19}$
& $20.94^{+12.80}_{-12.43}$ \\

\hline
\multicolumn{3}{c}{\textit{95\% upper credible limit}} \\
\hline

$\gamma$
& $\gamma_{95}<1.238$
& $\gamma_{95}<1.389$ \\

\hline\hline
\end{tabular}

\caption{
Constraints on the black-hole and dark-matter halo parameters obtained from the MCMC analysis using the Do et al.~\cite{Do19} and Gillessen et al.~\cite{Gillessen17ApJ}
datasets. The central values correspond to posterior medians with 68\% credible intervals. The one-sided 95\% upper credible limits on $\gamma$ are also shown.
}
\label{tab:mcmc_results}

\end{table*}

\begin{figure*}
\centering
\includegraphics[width=\textwidth]{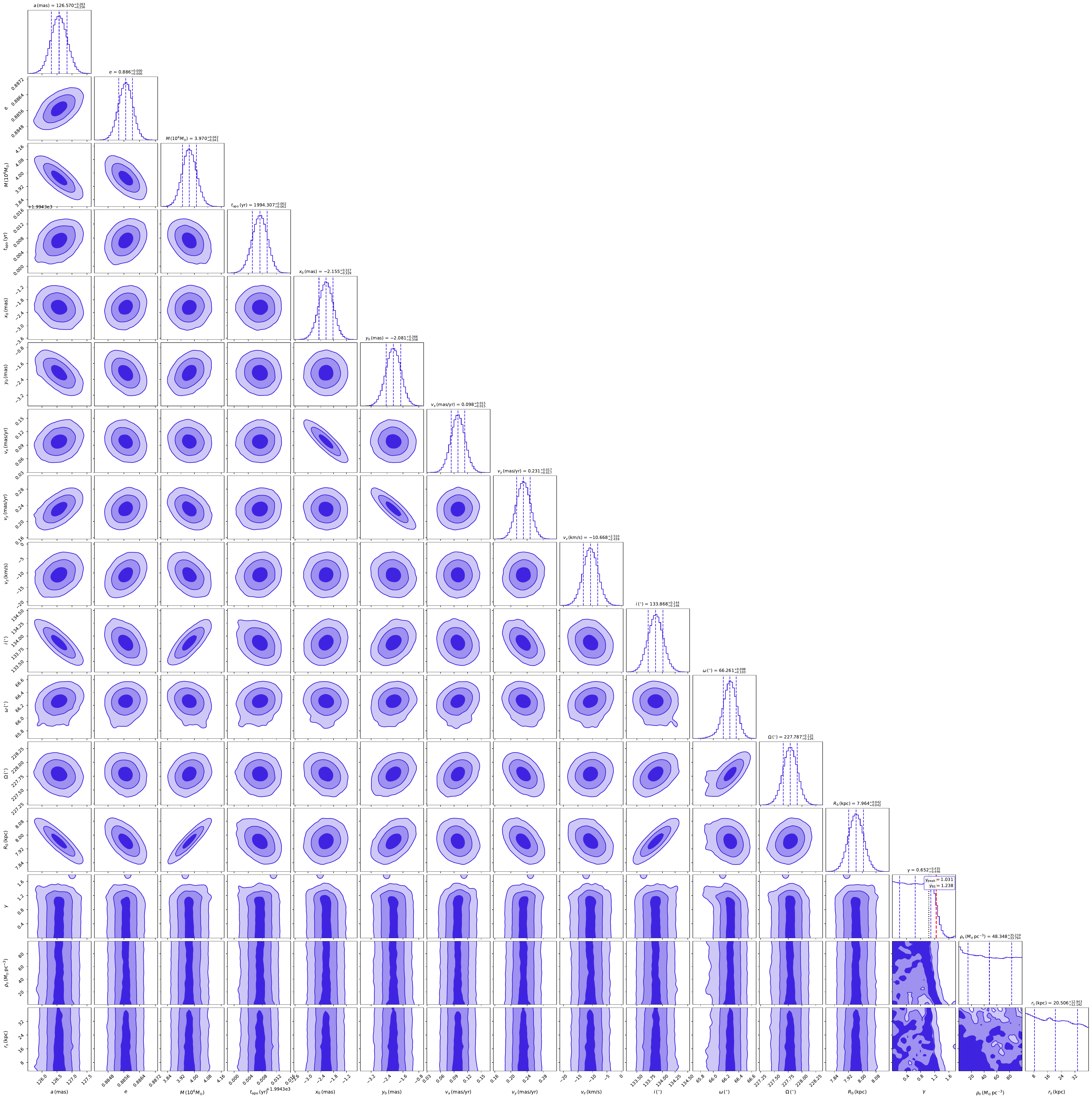}
\caption{\label{fig:posterior1} 
Posterior distributions of the model parameters obtained from the MCMC analysis for Dataset 1. The contours correspond to the 68\%, 95\%, and 99.7\% credible regions. The vertical dashed lines in the one-dimensional panels mark the posterior median and the 16th and 84th percentiles. In the $\gamma$ panel, the red dashed line marks the 95\% upper limit and ``peak'' denotes the position of the maximum of the posterior.
}
\end{figure*}

\begin{figure*}
\centering
\includegraphics[width=\textwidth]{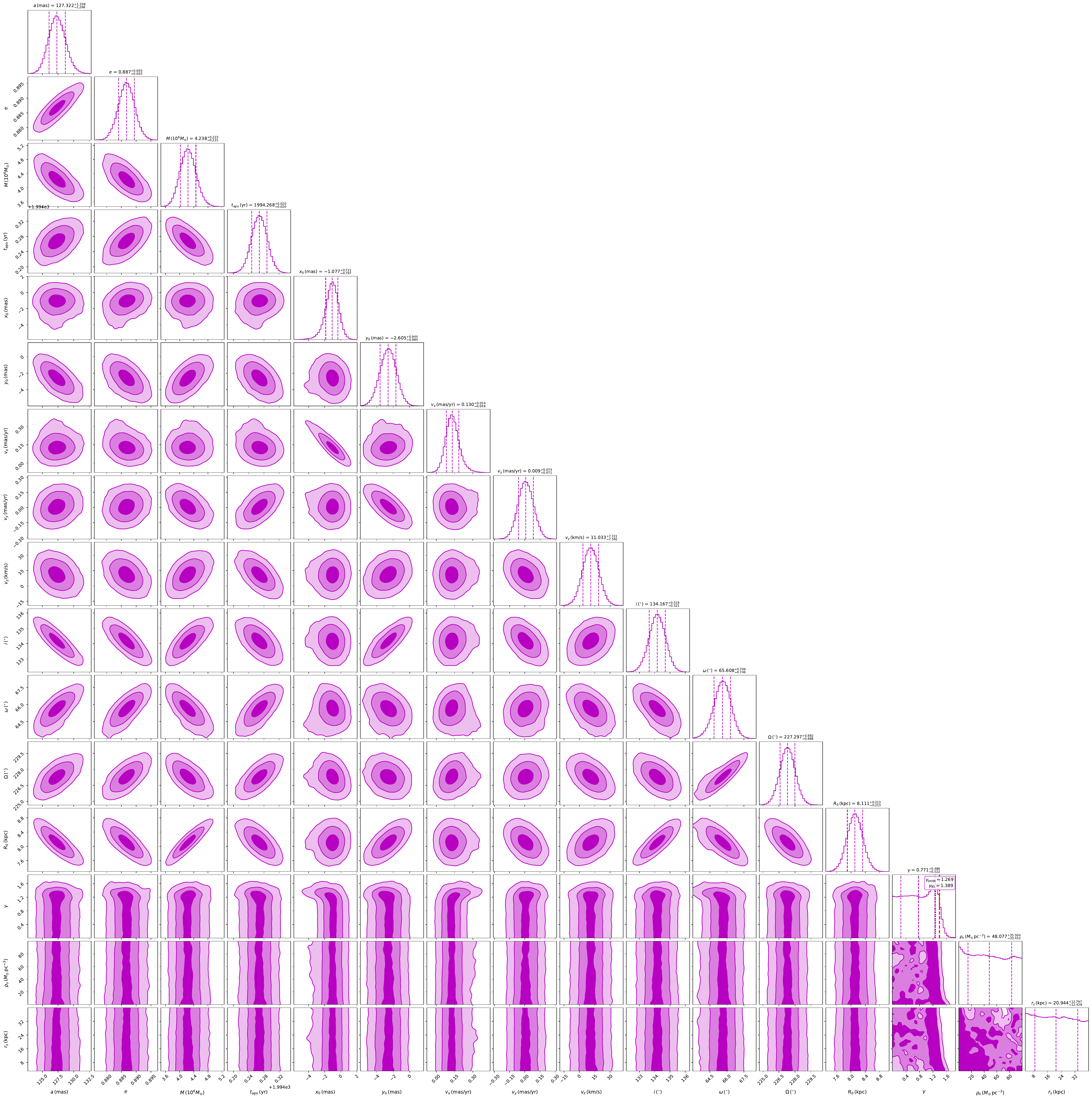}
\caption{\label{fig:posterior2} Posterior distributions of the model parameters obtained from the MCMC analysis for Dataset 2. The contours correspond to the 68\%, 95\%, and 99.7\% credible regions. The vertical dashed lines in the one-dimensional panels mark the posterior median and the 16th and 84th percentiles. In the $\gamma$ panel, the red dashed line marks the 95\% upper limit and ``peak'' denotes the position of the maximum of the posterior.}
\end{figure*}

Based on the above formulation and setup, we perform the MCMC analysis for both datasets. Figure~\ref{fig:orbit} shows the orbital trajectory and radial velocity of the S2 star evaluated at the posterior median parameters, together with the observational data. The corresponding posterior distributions for Dataset 1 and Dataset 2 are shown in Figs.~\ref{fig:posterior1} and \ref{fig:posterior2}, respectively, while the inferred parameter constraints are summarized in Table~\ref{tab:mcmc_results}.
For the dark-matter halo parameter $\gamma$, we obtain
$\gamma = 0.652^{+0.435}_{-0.446}$ for Dataset 1 and $\gamma = 0.771^{+0.486}_{-0.522}$ for Dataset 2, where the quoted uncertainties correspond to the 16th and 84th posterior percentiles (see, e.g., \cite{Hogg2018ApJS}). The corresponding one-sided
95\% upper credible limits are $\gamma < 1.238$ and
$\gamma < 1.389$ for Dataset 1 and Dataset 2, respectively.

The marginal posterior distributions of $\rho_s$ and $r_s$ show a strong degeneracy, which makes it difficult to independently constrain these two parameters. As a result, their posterior distributions remain sensitive to the adopted prior ranges. In contrast, the S2 orbital dynamics provide a useful constraint on the parameter $\gamma$, leading to meaningful upper limits. The posterior of
$\gamma$ is approximately flat at low $\gamma$ and shows a small peak around $\gamma\approx1.0$--$1.4$ before decreasing rapidly, but this does not constitute a clear preference for a particular value.

\begin{figure*}
	\centering
	\includegraphics[width=0.48\textwidth]{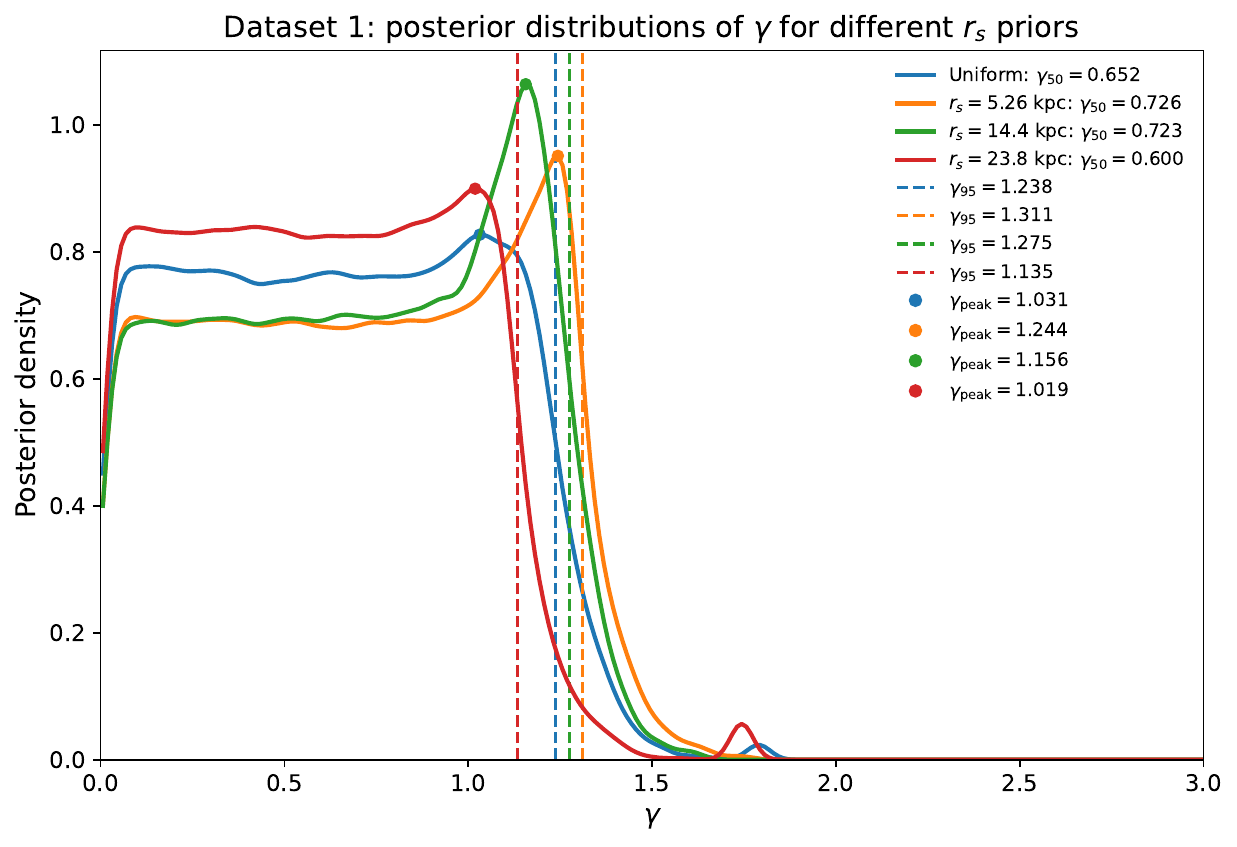}
    \includegraphics[width=0.48\textwidth]{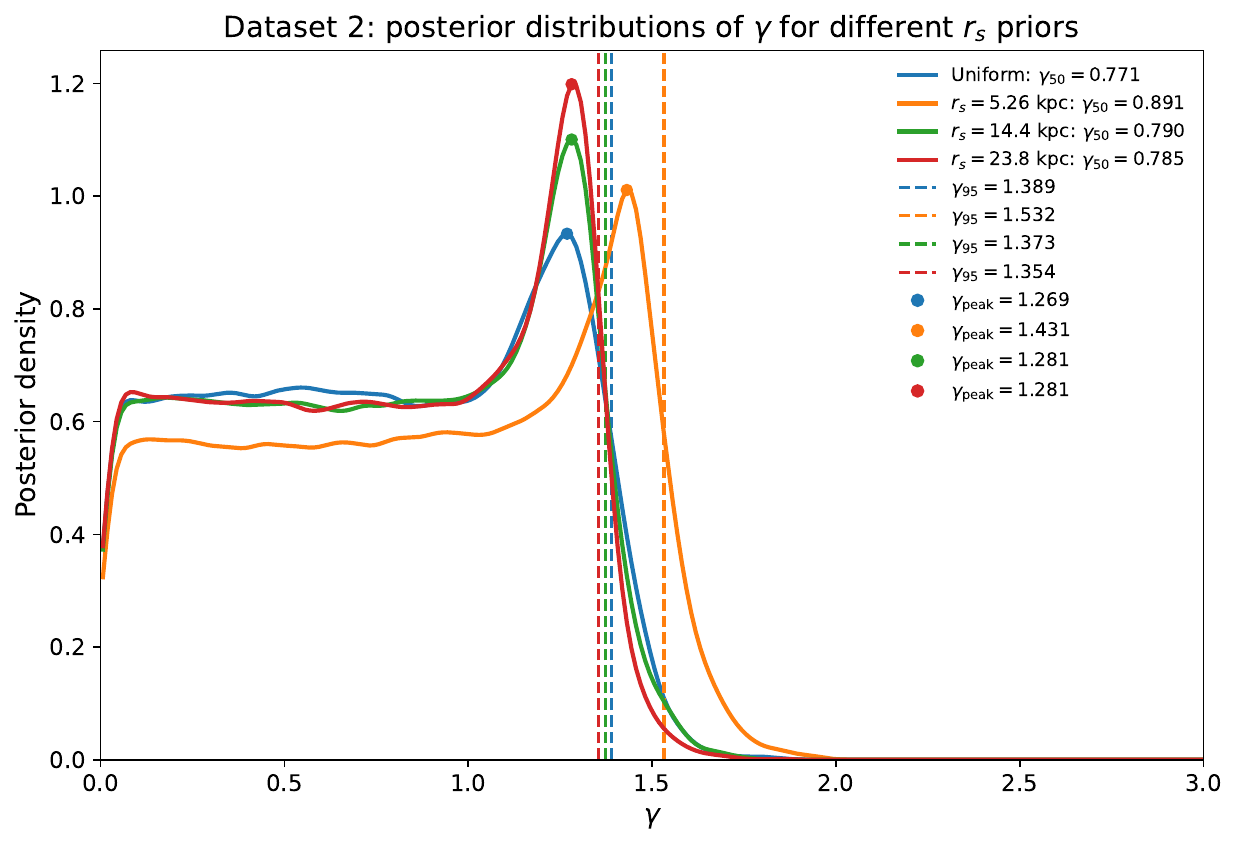}
	\caption{
    Marginalized posterior distributions of $\gamma$ for Dataset 1 (left) and Dataset 2 (right) under different priors on $r_s$: uniform $\mathcal{U}(3,40)$~kpc (blue) and split-Gaussian priors centered at 5.26 (orange), 14.4 (green) and 23.8~kpc (red). Vertical dashed lines mark the 95\% upper limits $\gamma_{95}$ and dots mark the posterior maxima.
    }  \label{fig:gammadensity}
\end{figure*}

To examine the sensitivity of the inferred constraints to the choice of the scale-radius prior, we additionally repeat the analysis using three observationally motivated split-Gaussian priors for $r_s$ \cite{Xiaowei_2024MNRAS, Cautun_2020MNRAS}, while keeping the priors of all other model parameters unchanged. The corresponding results are summarized in Table~\ref{tab:halo_constraints} and the marginalized posterior distributions of $\gamma$ are shown in Fig. \ref{fig:gammadensity}. For Dataset 1, the 95\% upper limit on $\gamma$ varies from $1.135$ to $1.311$, while for Dataset 2 it varies from $1.354$ to $1.532$. Thus, the constraint on $\gamma$ remains relatively stable under the different choices of the $r_s$ prior. The peak is higher for the Gaussian $r_s$ priors than for the uniform prior, as shown in Fig.~\ref{fig:gammadensity}. This is likely because the Gaussian priors limit the range of $r_s$, which reduces the degeneracy among $\gamma$, $\rho_s$ and $r_s$ and makes the peak more visible. It does not mean that the data prefer a particular value of $\gamma$.

The median halo parameters obtained with the broad prior on $\rho_s$ are $(\rho_s,r_s,\gamma)\simeq(48\,M_\odot\,{\rm pc}^{-3},\,20.5\,{\rm kpc},\,0.65)$. They imply a dark-matter density at the solar radius of $\rho_{\rm DM}(R_0)\sim30\,M_\odot\,{\rm pc}^{-3}$, far above the observed local value of $0.38\pm0.04$~GeV~cm$^{-3}$ ($0.0100\pm0.0011\,M_\odot\,{\rm pc}^{-3}$) \cite{McMillan2017}. To obtain a constraint consistent with this observation, we repeat the analysis with the halo normalized to the local density. Instead of $\rho_s$, we sample $\rho_{\rm loc}\equiv\rho_{\rm DM}(R_0)$ with a Gaussian prior given by this value and obtain $\rho_s$ from Eq.~\eqref{eq.density} at the sampled $R_0$,
\begin{equation}
\rho_s=\rho_{\rm loc}\left(\frac{R_0}{r_s}\right)^{\gamma}
\left(1+\frac{R_0}{r_s}\right)^{4-\gamma}.
\end{equation}
All other priors, including $\gamma\sim\mathcal{U}(0,3)$ and $r_s\sim\mathcal{U}(3,40)$~kpc, are unchanged. The resulting 95\% upper limits are $\gamma<1.949$ for Dataset 1 and $\gamma<1.931$ for Dataset 2 (Table~\ref{tab:halo_constraints}), with $\rho_s\simeq0.013\,M_\odot\,{\rm pc}^{-3}$. These limits are weaker than those obtained with the broad prior, because the DM halo normalized to the local density contains much less mass at the scale of the S2 orbit.


\begin{table*}[t]
\centering
\renewcommand{\arraystretch}{1.2}
\begin{tabular}{llcccc}
\hline\hline
Dataset & $r_s$ prior & $\gamma$ & $\gamma_{95}$ & $\rho_s\,[M_\odot\,\mathrm{pc}^{-3}]$ & $r_s\,[\mathrm{kpc}]$ \\
\hline
Dataset 1 & $\mathcal{U}(3,40)$ kpc & $0.652^{+0.435}_{-0.446}$ & $1.238$ & $48.35^{+35.22}_{-33.79}$ & $20.51^{+12.94}_{-12.19}$ \\
Dataset 1 & $5.26^{+0.15}_{-0.11}$ kpc & $0.726^{+0.464}_{-0.495}$ & $1.311$ & $48.80^{+34.95}_{-34.26}$ & $5.28^{+0.14}_{-0.12}$ \\
Dataset 1 & $14.4^{+4.5}_{-3.5}$ kpc & $0.723^{+0.432}_{-0.490}$ & $1.275$ & $48.81^{+34.30}_{-34.28}$ & $15.15^{+4.25}_{-3.87}$ \\
Dataset 1 & $23.8^{+8.1}_{-6.2}$ kpc & $0.600^{+0.405}_{-0.409}$ & $1.135$ & $47.49^{+35.72}_{-33.77}$ & $24.69^{+7.06}_{-6.73}$ \\
Dataset 2 & $\mathcal{U}(3,40)$ kpc & $0.771^{+0.486}_{-0.522}$ & $1.389$ & $48.08^{+35.50}_{-33.43}$ & $20.94^{+12.80}_{-12.43}$ \\
Dataset 2 & $5.26^{+0.15}_{-0.11}$ kpc & $0.891^{+0.520}_{-0.607}$ & $1.532$ & $47.87^{+34.88}_{-33.26}$ & $5.28^{+0.14}_{-0.12}$ \\
Dataset 2 & $14.4^{+4.5}_{-3.5}$ kpc & $0.790^{+0.473}_{-0.540}$ & $1.373$ & $48.95^{+34.73}_{-34.36}$ & $14.96^{+4.20}_{-3.82}$ \\
Dataset 2 & $23.8^{+8.1}_{-6.2}$ kpc & $0.785^{+0.475}_{-0.538}$ & $1.354$ & $50.01^{+34.47}_{-34.28}$ & $24.38^{+7.17}_{-6.53}$ \\
\hline
& &$\rho_{\rm loc}=0.0100\pm0.0011\,M_\odot\,{\rm pc}^{-3}$ \citep{McMillan2017}\\
\hline
Dataset 1 & $\mathcal{U}(3,40)$~kpc & $1.068^{+0.720}_{-0.730}$ & 1.949 & $0.013^{+0.050}_{-0.010}$ & $21.46^{+12.62}_{-12.69}$ \\
Dataset 2 & $\mathcal{U}(3,40)$~kpc & $1.027^{+0.697}_{-0.700}$ & 1.931 & $0.013^{+0.056}_{-0.010}$ & $21.49^{+12.54}_{-12.51}$ \\
\hline\hline
\end{tabular}
\caption{Posterior constraints on the generalized Dehnen-halo parameters obtained from the two S2 datasets under different priors on the scale radius $r_s$ {and on the halo normalization}. The quoted uncertainties correspond to the 16th and 84th posterior percentiles around the median, while $\gamma_{95}$ denotes the one-sided 95\% upper credible limit. Here, $\mathcal{U}(3,40)$ denotes a uniform prior, while the priors $\mu^{+\sigma_+}_{-\sigma_-}$ are split-Gaussian distributions. The upper rows use $\rho_s\sim\mathcal{U}(0,100)\,M_\odot\,{\rm pc}^{-3}$ with different priors on $r_s$. The two lower rows instead use a Gaussian prior on the local dark-matter density, $\rho_{\rm loc}=0.0100\pm0.0011\,M_\odot\,{\rm pc}^{-3}$ \citep{McMillan2017}, from which $\rho_s$ is derived.
}
\label{tab:halo_constraints}
\end{table*}

\section{\label{conclusion} Conclusion}

In this paper, we studied the dynamics of massive particles around a Schwarzschild-like BH surrounded by a Dehnen-type DM halo characterized by the density profile $(1,4,\gamma)$, considering the solution as a possible description of Sgr A$^{\star}$ at the center of the Milky Way that serves as a valuable laboratory for probing BH-DM halo system properties. We derived the equations of motion and obtained the pericenter shift over one orbital period. With this we further presented a comprehensive investigation of Schwarzschild-like BH within a Dehnen-type DM halo, exploring parameter estimation using the orbit of the S2 star around Sgr A$^{\star}$. We developed a complete model that connects theoretical predictions with current observational constraints from Dataset~1 of Do et al.~\cite{Do19}, and Dataset~2 of Gillessen et al.~\cite{Gillessen17ApJ}.

Applying this framework to the orbit of the S2 star around Sgr A$^\star$, we investigated the constraints on the parameters of the generalized Dehnen-type dark-matter halo using two independent observational datasets from Do et al.~\cite{Do19} and Gillessen et al.~\cite{Gillessen17ApJ}. 
The MCMC analysis provides constraints on the halo parameter $\gamma$. The posterior medians with 68\% credible intervals are $\gamma=0.652^{+0.435}_{-0.446}$ and $\gamma=0.771^{+0.486}_{-0.522}$ for Dataset~1 and Dataset~2, respectively. The corresponding one-sided 95\% upper credible limits are $\gamma<1.238$ and $\gamma<1.389$.

When we normalized the DM halo to the observed local dark-matter density at solar distance, $\rho_{\rm DM}(8\, \rm kpc)\simeq0.01\,M_\odot\,{\rm pc}^{-3}$ \cite{McMillan2017},
the 95\% upper limits become $\gamma<1.949$ and $\gamma<1.931$ for Dataset 1
and Dataset 2, respectively.
Although the S2 orbital data provide a useful constraint on the halo parameter $\gamma$, the parameters $\rho_s$ and $r_s$ exhibit a strong degeneracy and their posterior distributions remain sensitive to the adopted prior ranges.
Therefore, the S2 orbital data alone do not provide independent constraints on the halo density and scale radius.

\section*{Acknowledgements}{
We are grateful to the anonymous referee for the constructive comments and suggestions that helped us improve the clarity and quality of this work.
The research is supported by the National Natural Science Foundation of China under Grant No. W2433018. 
}

\bibliographystyle{apsrev4-1}
\bibliography{gravreferences,ref}



\end{document}